\documentclass[%
 aip,
 amsmath,amssymb,
 reprint,%
]{revtex4-2}

\usepackage{graphicx}
\usepackage{dcolumn}
\usepackage{bm}
\usepackage[utf8]{inputenc}
\usepackage[T1]{fontenc}
\usepackage{mathptmx}
\usepackage{etoolbox}

\usepackage[mathlines]{lineno}
\usepackage[per-mode=symbol]{siunitx}
\usepackage{hyperref}
\hypersetup{
  colorlinks=true,
  linkcolor=blue,
  citecolor=blue,
  urlcolor=blue
}

\makeatletter
\def\@email#1#2{
 \endgroup
 \patchcmd{\titleblock@produce}
  {\frontmatter@RRAPformat}
  {\frontmatter@RRAPformat{\produce@RRAP{*#1\href{mailto:#2}{#2}}}\frontmatter@RRAPformat}
  {}{}
}
\makeatother

\graphicspath{{Figures/}}
\begin{document}
\preprint{AIP/123-QED}

\title[]
{Absorption imaging of quantum gases near surfaces using incoherent light}

\author{Julia Fekete}
\author{Poppy Joshi}
\author{Peter Kr\"uger}
\altaffiliation[Also at ]{Physikalisch-Technische Bundesanstalt, 10587 Berlin, Germany}
\author{Fedja Oru\v cevi\'c}

\affiliation{Department of Physics and Astronomy, University of Sussex, Brighton BN1 9QH, United Kingdom}
\email{F.Orucevic@sussex.ac.uk}

\begin{abstract}

We introduce an absorption imaging technique for ultracold gases that suppresses interference fringes and coherence-induced artifacts by reducing the transverse spatial coherence of the imaging light. The method preserves the narrow spectral bandwidth required for resonant absorption imaging and is implemented as a modular extension to standard imaging setups using a rotating diffuser. We demonstrate tunability of the illumination light's coherence without modifying the imaging optics. Using this approach, we achieve reliable imaging of ultracold atomic clouds in micron-scale proximity to complex surfaces, where standing waves, edge diffraction, and speckle severely limit conventional absorption imaging.

\end{abstract}

\maketitle

A wide range of physical systems involve atoms or other particles located near surfaces, where imaging serves as the primary tool for characterizing their properties. We consider
 ultracold neutral atomic gases trapped and manipulated by chips or other planar structures. They provide a versatile platform to 
(i) probe atom-surface interactions \cite{Schneeweiss2012,Hummer2021,Laliotis2021}, local magnetic fields \cite{Wildermuth2005,Yang2020}, and current flow irregularities \cite{Yang2017,Fekete2024}; 
(ii) coherently manipulate atomic states for atom interferometry \cite{Wang2005};  and 
(iii) study quantum many-body systems \cite{Hofferberth2007}.

Absorption imaging using resonant or near-resonant light remains the most widely used technique for the quantitative observation of the spatial distributions of quantum gases. The relevant atomic transitions have linewidths on the order of megahertz or below. As a result, the imaging light is typically derived from a highly coherent Gaussian laser beam with coherence times of microseconds and longitudinal coherence lengths of hundreds of meters. 
As a consequence, interference effects are ubiquitous and often arise as a combination of 
(i) standing waves formed by reflection on a nearby surface; 
(ii) diffraction at trapping or mounting structure edges; 
and (iii) speckles or more regular fringes formed on optical elements due to contaminants. 
A standard procedure to mitigate illumination-related artifacts involves acquiring three consecutive images: a shadow image of the atomic distribution, a reference light image without atoms, and a reference dark image, with millisecond time delays between them.
While this approach effectively compensates for static illumination inhomogeneities, it does not perform well in the presence of high-contrast interference fringes or time-dependent phase modulations, such as those induced by air turbulence. Such modulations are imprinted onto the resulting optical density.

Following the comprehensive study of absorption imaging of quantum gases near atom chips where standing waves are formed \cite{Smith2011}, various post-processing algorithms have been developed to remove interference-related artifacts \cite{Niu2018,Xiong2020,Li2007}. Some of these rely on strong assumptions about the underlying atomic distribution (for example Thomas-Fermi or Gaussian), and by construction they cannot recover information about the atoms where the illumination vanishes. Finite-sized and partially reflective components further complicate the illumination through multiple reflections, waveguiding within substrates, edge diffraction, and scattering. This can result in complex intensity patterns, sensitive to the beam alignment. In some cases, atoms near a surface lie in a region where adequate illumination cannot be achieved. 

These limitations can be mitigated by using spatially incoherent illumination for absorption imaging, hereafter referred to as \textit{incoherent imaging}. 
Techniques for reducing the spatial coherence of a light source have been successfully implemented for various applications by diverse methods including the use of diffusers with incidence angle modulation \cite{Wei2019,Akram2010}, moving \cite{Kubota2010} or rotating \cite{Stangner2017,Wei2019} diffusers, random lasers \cite{Redding2012} or intra-cavity masks \cite{Chriki2015}. 
However, implementing incoherent imaging for fragile ultracold clouds remains challenging, as it requires (i) resonant or near-resonant illumination within the narrow atomic linewidth and (ii) microsecond-scale exposure times.

In this paper, we introduce a method for incoherent imaging of ultracold gases based on suppressing the transverse spatial coherence of the narrow-band imaging beam by a rotating diffuser. The technique can be implemented as a modular extension to existing absorption imaging setups. 
Using this approach, we overcome limitations associated with coherence-induced effects and enable reliable imaging of atoms in regions dominated by edge diffraction or speckles. 
 
We further report the observation of a phenomenon producing artifacts in the spatial distribution of quantum gases imaged after short time-of-flight (TOF) expansion. The artifacts appear in the two-dimensional density distribution and are carried over to the extracted longitudinal line density. By comparing measurements performed with coherent and incoherent illumination, we identify their origin as coherence-related imaging effects. The ability to control the spatial coherence of the imaging light provides a powerful diagnostic tool for distinguishing genuine physical features from coherence-induced artifacts and noise.

In our approach to reducing the coherence of the imaging light, it is essential that the spectral properties and thereby the coherence time, which is inversely proportional to the spectral bandwidth, are not altered. We reduce the spatial coherence of the light instead, by introducing spatially random phase shifts via a scattering medium, which modulates the wavefront of the imaging beam. The resulting speckle field still constitutes a coherent noise pattern, but temporal variation of the field leads to reduction of the transverse spatial coherence. 
The condition for the light field to become spatially incoherent is linked to a sufficiently large displacement of the random pattern during the observation time. In practice this means that the imaging exposure time needs to be much longer than the correlation time, $t_c$, associated with the transverse correlation length. 
The ideal scattering medium to create random phase-modulated patterns is an optical diffuser with high transmittance. 
We induce temporal variation by rotating the diffuser (at frequency $f$), on which the laser beam is incident at a large radial distance, $R$, from the rotation axis. The speckle size grows with the tangential velocity $v=2\pi f R$, while the speckle contrast decreases, ultimately resulting in homogeneous illumination. To characterize the relevant parameters in our experiment, we analyzed the speckle pattern generated by a collimated Gaussian beam of \qty{2.1}{\mm} diameter on the static diffuser. A correlation length of $l_c \approx \qty{40}{\um}$ and correlation time of $t_c = l_c /v \approx \qty{1}{\us}$ are estimated for $v = \qty{42}{\meter\per\second}$. Typical exposure times in absorption imaging are tens of microseconds, far exceeding $t_c$, while specifically in our experiment, we found that exposure times above \qty{12}{\us} sufficiently average over speckle realizations and result in a smooth illumination profile.

\begin{figure}
\includegraphics[width =1\linewidth]{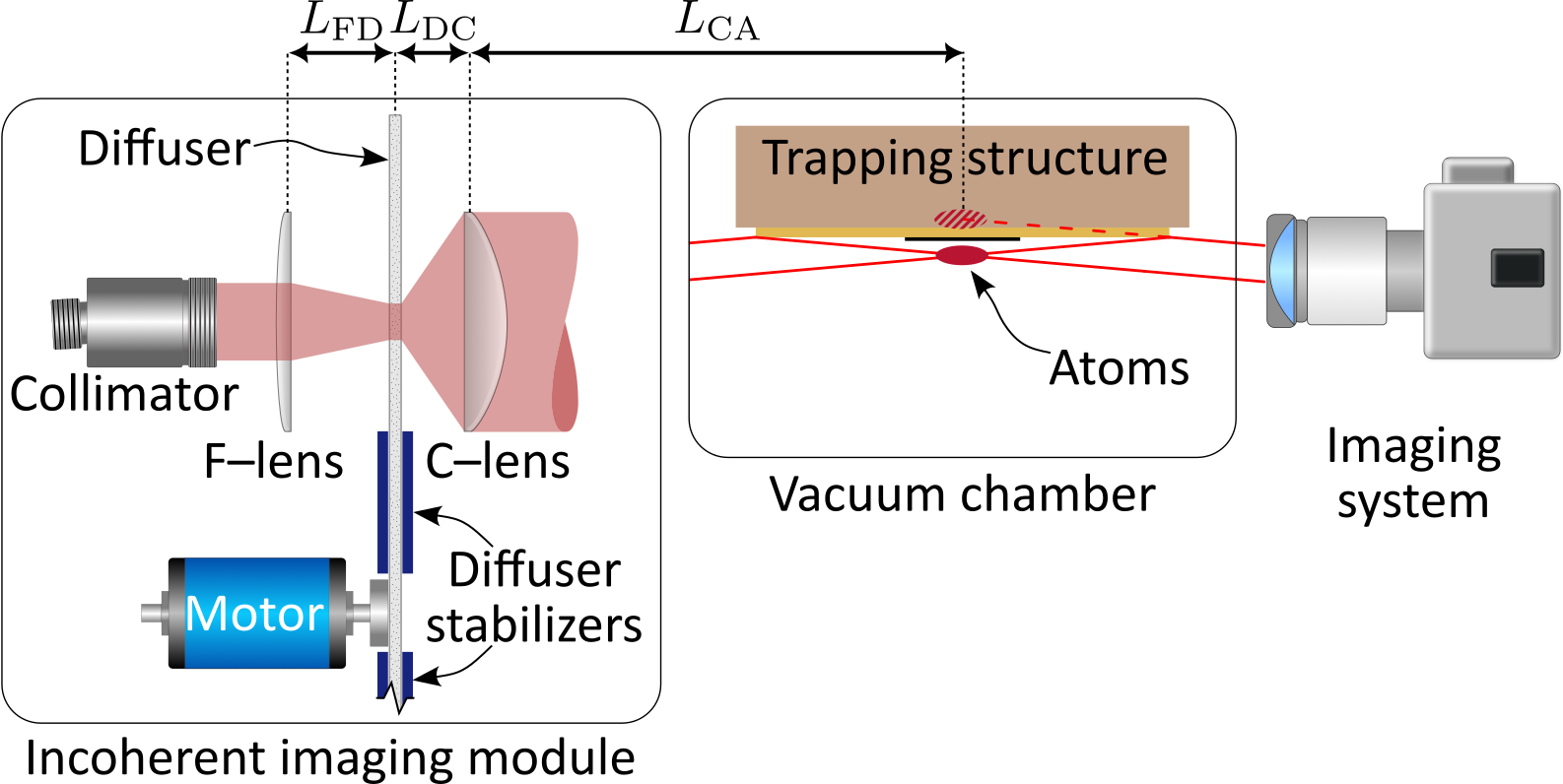}
\caption{\label{fig:scheme} Experimental scheme including the incoherent imaging module, the vacuum chamber with atoms below the planar trapping structure, to which a sample on a substrate is attached, and imaging system with the camera. The optional F-lens can be moved to tune the coherence of the light. Without the incoherent imaging module inserted, a large collimated beam provides coherent illumination to the atoms following the standard absorption imaging scheme. Red lines represent rays of particular interest for the image formation of atoms and their mirrored (dashed line) image.}
\end{figure}

\begin{figure*}
\includegraphics[width =1\linewidth]{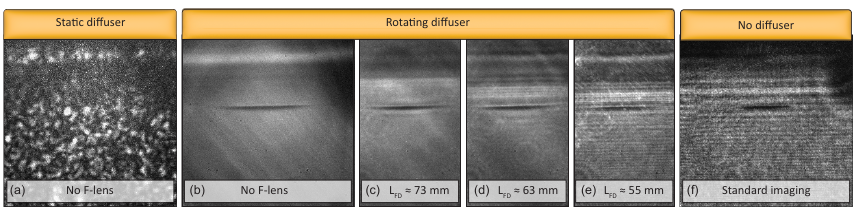}
\caption{\label{fig:rawImgs}Raw images with an ultracold cloud placed at \qty{430}{\um} distance from the surface. (a) Speckle pattern obtained with the incoherent imaging setup with a static diffuser and C-lens, without F-lens. (b) Image taken with the same setup, but the diffuser rotating at  \qty{90}{\hertz}, as used throughout the paper. (c-e) Images demonstrating the tunability of coherence by moving the additional F-lens along the optical axis. Diffuser and C-lens are unchanged relative to (b). (f) Standard absorption imaging scheme with the incoherent imaging module fully removed.}
\end{figure*}

In a typical absorption imaging scheme, a large diameter collimated beam illuminates the atoms and is imaged onto a camera (see Fig.~\ref{fig:scheme}). Without modifying the imaging optics or camera alignment, we insert a modular unit into the optical path at the entrance of the vacuum system, to modify the spatial coherence of the imaging light. In this module, a beam with up to \qty{60}{\milli\watt} power from a fiber collimator (Thorlabs F220APC-780) is incident on a diffuse sheet of tracing paper. 
The optical transmittance of the diffuser is \qty{69}{\percent} and the measured grain size distribution reaches up to \qty{15}{\um}. 
The diffuser sheet, cut into a disk shape, is mounted concentric with a drone motor (LDF Power MT2213-920 KV), rotating at $f \approx \qty{90}{\Hz}$ to avoid mechanical instabilities occurring at higher speeds. The beam intersects the diffuser near its outer edge, at a radial position of $R \approx \qty{75}{\mm}$. To redirect a large fraction of the light scattered by the diffuser, we place a condenser lens (C-lens, Thorlabs ACL50832U-B) before the vacuum chamber. The distance between C-lens and the atomic cloud is $L_{\mathrm{CA}}\approx \qty{500}{mm}$, determined by the size of the vacuum chamber. The distance between the diffuser and C-lens of $L_{\mathrm{DC}}\approx \qty{16}{\mm}$ was found to maximize the contrast in tests carried out with a calibration target. Figs.~\ref{fig:rawImgs} (a,b) show the raw images recorded in this configuration with static and rotating diffusers, respectively, when atoms are trapped \qty{430}{\um} below the surface.

The beam size on the sheet is an important control parameter for the imaging technique. As point-like sources yield large transverse coherence, focusing the incoming beam onto the diffuser can recover the coherence lost during scattering. 
To explore this, we add a focusing lens (F-lens, Thorlabs LA1145-B) before the diffuser and vary their distance $L_{\mathrm{FD}}$ to tune the degree of coherence. 
We observe a clear transition from incoherent ($L_{\mathrm{FD}}\approx \qty{73}{\mm}$, beam diameter $d_0\approx 1$~mm at the diffuser), through partially coherent ($L_{\mathrm{FD}}\approx \qty{63}{\mm}$, $d_0\approx 0.5$~mm) back to coherent light ($L_{\mathrm{FD}}\approx \qty{55}{\mm}$, $d_0\approx \qty{70}{\um}$), exhibiting interference fringes. Figs.~\ref{fig:rawImgs} (c-e) show representative images. 
The ratio of the beam area to the area associated with the correlation length determines the number of mutually incoherent modes or effective scatterers participating in the dynamical coherence reduction process. 
Our observation is in agreement with predictions describing the relation between the level of coherence and the number of incoherent modes \cite{Goodman2020}. 
Moreover, we can qualitatively reproduce the interference structure formed in the standard absorption imaging configuration, where the coherent imaging beam is expanded to \qty{30}{mm} diameter (see Fig.~\ref{fig:rawImgs} (f)). Residual quantitative deviations between Figs.~\ref{fig:rawImgs} (e) and (f) stem from the presence of different optical surfaces involved and the sensitivity to alignment in the two configurations.

\begin{figure*}
\includegraphics[width =1\linewidth]{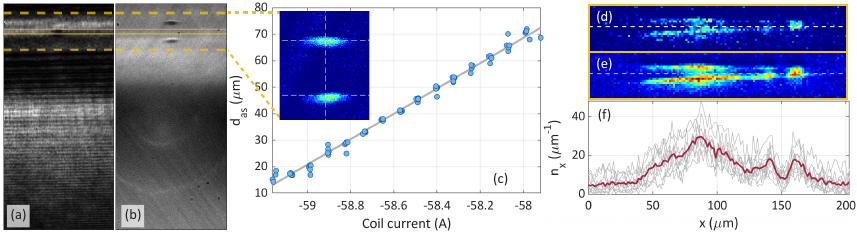}
\caption{\label{fig:ASdistance} 
Raw images of atoms in the near vicinity of the surface, using (a) coherent and (b) incoherent light. A region centered on the surface location is indicated with yellow solid (dashed) line with vertical extent of 34 (270) \unit{\um}. (c: inset) Atomic distribution recorded with incoherent light when atoms are placed at $d_{\textrm{as}} \approx \qty{62}{\um}$ height below the surface, appearing as two clouds separated vertically by $2d_{\textrm{as}}$ distance. Dashed lines indicate center-of-mass positions extracted from fits to the vertical and horizontal profiles. (c) Atom-surface distance $d_{\textrm{as}}$ calibration to trapping current, used for moving the atoms: current in the Helmholtz coil pair provides bias field for a Z-wire trap where the Z-wire current is kept constant at $85.0 \pm \qty{0.03}{\ampere}$. Each data point is obtained using Gaussian fits on the atomic clouds and their mirror images while the currents were recorded. Closer to the surface, the two clouds merge, and the multi-Gaussian fit becomes unreliable. 
 (d-e) \qty{34}{\um} height section of ACD images of atoms trapped within \qty{6}{\um} from the surface. (d) is a single and (e) is average of 10 images taken with incoherent imaging. Dashed lines indicate the surface. (f) Linear density profiles calculated for data in (d-e), with single profiles in gray, mean profile in dark red. }
\end{figure*}

Absorption imaging of atoms near a surface is commonly performed by directing the imaging beam from below at grazing incidence relative to the surface \cite{Smith2011}. This geometry is essential for determining the atom-surface distance, $d_{\textrm{as}}$, via the separation of the real and mirror images of the atoms. In our experiment, the atoms are placed near a quartz substrate that carries an opaque nano-structured sample under study in our quantum gas microscope, setup described elsewhere \cite{Fekete2024}. Unlike reflection from an atom chip surface, this sample-substrate structure generates more complex coherent light illumination patterns that can prevent simultaneous imaging of the cloud and its mirror image (see Fig.~\ref{fig:ASdistance}~(a)). This precludes a reliable trapping current calibration \cite{Smith2011,Yang2017} of $d_{\textrm{as}}$ in the sub-\qty{100}{\um} range. 
By contrast, with spatially incoherent light we obtain a uniformly illuminated region that extends well beyond the range accessible under coherent imaging (Fig.~\ref{fig:ASdistance}~(b)). This configuration enables a robust trapping current calibration (Fig.~\ref{fig:ASdistance}~(c)). The inset shows an example atomic column density (ACD) image displaying atoms and their mirror image separated by \qty{124}{\um}.

To investigate the atomic response to surface effects such as variations of the atomic distribution when exposed to magnetic fields from nearby current-carrying structures, we trap atoms in close proximity to our sample, and image them with the real and mirror images of the cloud both being visible.  Fig.~\ref{fig:ASdistance}~(d) shows an example ACD image of an atomic cloud trapped within \qty{6}{\um} of the surface, obtained using incoherent imaging. Fig.~\ref{fig:ASdistance}~(e) shows the average of ten such images, and the corresponding vertically integrated profiles are shown in Fig.~\ref{fig:ASdistance}~(f). Individual profiles (gray lines) exhibit constant baseline offsets relative to the mean profile (dark red line), linked to regions with varying illumination levels. We attribute the presence of these regions to a deviation of the diffuser sheet from an ideal flat surface, which remains in spite of the firm mechanical support to the non-rigid diffuser sheet provided by thin 3D-printed disks (of 70~mm radii) from both sides. 
The constant offsets can be readily subtracted. For larger atomic distributions a linear background subtraction might be needed but does not limit the extraction of line densities. Our trap simulations indicate a tilt toward the surface of approximately \(1^\circ\), matching our observations in Figs.~\ref{fig:ASdistance}~(d-e). From these data, we extract $d_{\textrm{as}} \leq \qty{5.8}{\um}$ (closer to the surface toward the right end of the figure). According to the same simulation, the radial size of the elongated quantum gas is well below the near-diffraction-limited optical resolution of our imaging system that we estimated from an independent measurement, and the vertical cloud size of \qty{3.5}{\um} in Fig.~\ref{fig:ASdistance}~(e) is in good agreement with that.

\begin{figure}
\includegraphics[width =0.8\linewidth]{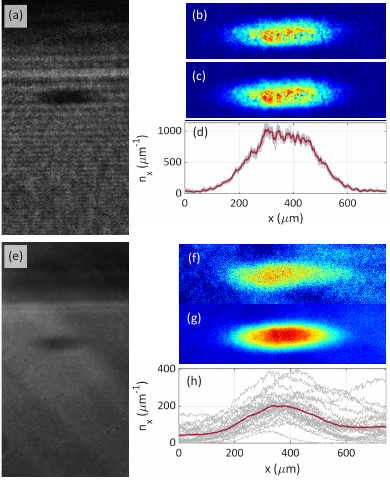}
\caption{\label{fig:thermalcloud} (a,e) Raw images, (b,f) single and (c,g) averaged ACD of 25 repeats, and (d,h) the corresponding linear density profiles, for coherent (a-d) and incoherent (e-h) light. The data were taken of a thermal cloud after \qty{1}{\ms} TOF.}
\end{figure}

To demonstrate the advantages of incoherent imaging in a different parameter range, we prepared thermal clouds of approximately \qty{5}{\micro\kelvin} temperature at $d_{\textrm{as}} \approx \qty{430}{\um}$, and imaged them after \qty{1}{\ms} TOF. With coherent imaging light at grazing incidence, pronounced interference fringes are visible in the raw images (Fig.~\ref{fig:thermalcloud}~(a)), which arise as a result of standing wave formation upon reflection from the surface, combined with edge diffraction from the mounting structure. As expected, these fringes are absent when using incoherent light (Fig.~\ref{fig:thermalcloud}~(e)). 
The high-contrast fringes in Fig.~\ref{fig:thermalcloud}~(a) appear equally in  the shadow and reference light frames and are therefore largely removed in the ACD reconstruction. This cancelation is effective where the two frames have similar intensity, i.e. away from the atomic signal. In the region containing atoms, however, the shadow and reference intensities differ strongly, and a speckle-like background pattern remains. Moreover, because this pattern is essentially identical from shot to shot, averaging multiple ACD images does not remove it, and can be misinterpreted as physical density modulation. Figs.~\ref{fig:thermalcloud}~(b-d) show single and averaged ACD images of 25 repeats, and corresponding linear density profiles, where the modulation is carried over, respectively. 
Note that dynamic speckles, formed for example by particles in turbulent air, would average out. Incoherent imaging is free from speckle-induced artifacts, yielding the expected smooth Gaussian density profile after averaging (see ACD images in Figs.~\ref{fig:thermalcloud}~(f,g)).

To address the limitations of incoherent imaging, we analyze the linear density profiles in Fig.~\ref{fig:thermalcloud}. 
The uneven background on the single-shot profiles originates from the mechanical instability of the rotating diffuser, as mentioned above, and can be identified in Fig.~\ref{fig:thermalcloud}~(e) as diagonal sections with varying illumination level, on spatial scales comparable to the cloud size. These residual effects can be reduced by an appropriate choice of the orientation of the features determined by the beam position on the rotating diffuser, and by subtracting the linear baseline from the averaged profiles. 
According to coherent imaging theory, the degree of spatial coherence affects the image transfer function and contrast. In comparison to coherent light, better resolution can be achieved in principle, at the expense of a reduction in image contrast with incoherent light. Consistent with this prediction, we observe reduced atomic densities in Fig.~\ref{fig:thermalcloud}~(h), compared to the coherent case in Fig.~\ref{fig:thermalcloud}~(d). However, this can be corrected by calibrating the atomic densities with coherent imaging.

\begin{figure*}
\includegraphics[width =1\linewidth]{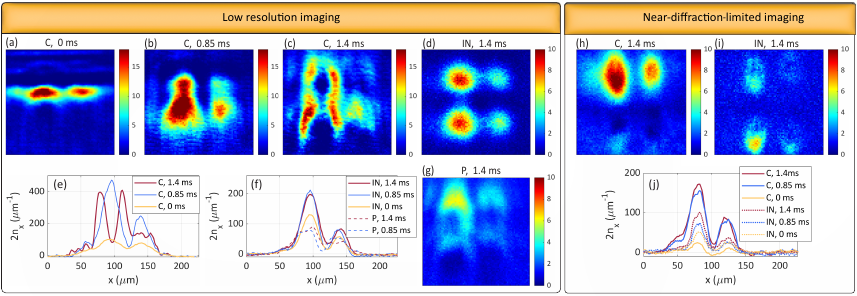}
\caption{\label{fig:tofscan} (a-g) ACD images and resulting linear densities for data taken with an imaging system comprising optical aberrations. (a-c) Atomic distributions taken in trap and after 0.85 and \qty{1.4}{\ms} TOF, respectively, using coherent light. Pronounced imaging aberration effects can be observed as fringes above the in-situ cloud. (d) Linear density profiles from images (a-c) indicate almost complementary distributions for the two distributions measured after a TOF. (e) Linear densities for incoherent and partially coherent light. Aberrations for data taken with incoherent light cause suboptimal resolution but no visible artifacts. (f-g) ACD images taken after \qty{1.4}{\ms} TOF with incoherent and partially coherent light. 
(h-j) Data taken with a near-diffraction-limited imaging system. (h-i) ACD images taken after \qty{1.4}{\ms} TOF for coherent and incoherent light. (j) Linear densities for various TOF using coherent and incoherent imaging.}
\end{figure*}

As a final example, we demonstrate the benefit of controlled spatial coherence to distinguish genuine atomic density features from coherence-induced imaging artifacts. 
We insert the coherence reduction module into the existing imaging setup, and present a situation where the role of coherence itself is tested. 
We prepare atoms in a magnetic trap and image the distribution in-situ and after short TOF expansion, using an imaging system that exhibits optical aberrations.  
Figs.~\ref{fig:tofscan}~(a-c) show distributions taken in-situ, after 0.85 and \qty{1.4}{\ms} TOF, respectively, using coherent illumination (labeled as C). The corresponding linear densities (Fig.~\ref{fig:tofscan}~(d)) reveal nearly complementary profiles at the two expansion times, indicating an imaging artifact. Reducing the imaging beam’s spatial coherence suppresses this behavior: the artifact is eliminated with incoherent (labeled as IN) illumination and partially suppressed with partially coherent (P) illumination (see images taken after \qty{1.4}{\ms} TOF in Figs.~\ref{fig:tofscan}~(f,g) and linear densities in Fig.~\ref{fig:tofscan}~(e)). 
The systematic suppression of the artifact as the coherence is reduced points to coherence-induced imaging effects.

We repeated the measurements using a near-diffraction-limited imaging system with a resolution of approximately \qty{3.5}{\um}. In this case, the extracted linear densities remain consistent for all levels of coherence, indicating that optical aberrations are further contributors to the observed anomaly in the atomic distributions. Fig.~\ref{fig:tofscan}~(j) displays profiles with very similar shapes at various TOF values and coherence levels. The overall atom number densities and their scaling with TOF are consistent with atom number estimates in the presence of fields decaying upon trap relaxation. The remaining non-trivial features such as the asymmetry of the mirrored images at \qty{1.4}{\ms} TOF, shown in Figs.~\ref{fig:tofscan}~(h,i), are attributed to residual optical aberration, although phase shifts upon reflection may also play a role. 
We conclude that the behavior presented in Fig.~\ref{fig:tofscan} arises from an interplay between optical aberrations and coherence phenomena. A deeper understanding of the underlying mechanism is beyond the scope of this paper. 

In conclusion, we have introduced a simple and robust method to suppress the transverse spatial coherence of the narrow-band imaging light used in cold atom experiments, thereby eliminating interference effects such as speckles, standing waves, and edge-diffraction patterns. We demonstrated reliable absorption imaging of ultracold atomic gases in spatial regions where conventional coherent imaging fails or produces artifacts.
The resulting extended field-of-view in close proximity to surfaces enables continuous and accurate determination of atom–surface distances, even in the presence of non-trivial optical and surface properties. This significantly improves the calibration of experimental control parameters, such as trapping wire currents and bias fields. 
In addition, we showed that implementing coherence control as a modular extension of existing imaging setups provides a powerful diagnostic tool. By comparing measurements performed with coherent, partially coherent, and incoherent illumination, we identified an interplay between spatial coherence and optical aberrations that can give rise to reproducible but non-physical features in atomic density distributions. Owing to its simplicity and broad applicability, the method can be readily integrated into experiments involving fragile atomic systems or imaging near surfaces. This will open the pathway to new and improved applications in the areas of cavity quantum electro-dynamics, quantum control, cold atoms near fibers, optical sensing and quantum technologies in general.

\begin{acknowledgments}
We wish to acknowledge the support of EPSRC (EP/Z535503/1). We thank Alice King for the help in fabricating the nano-structured sample.
\end{acknowledgments}

\section*{Author Declarations}
\textbf{Conflict of Interest}
The authors have no conflicts to disclose.

\section*{Author Contributions}
\textbf{Julia Fekete:} conceptualization (lead); investigation (lead); formal analysis (equal); writing – original draft (lead); visualization (equal). 
\textbf{Poppy Joshi:} investigation (equal); formal analysis (lead); writing – review and editing (equal); visualization (equal). 
\textbf{Peter Kr\"uger:} conceptualization (equal); writing – review and editing (equal).
\textbf{Fedja Oru\v cevi\'c:} conceptualization (equal); writing – review and editing (equal); visualization (equal).

\section*{Data Availability}
Data supporting this study are openly available from Figshare of the University of Sussex at DOI

\section*{References}
\bibliography{references}

\end{document}